\documentclass[aps,prb,superscriptaddress,reprint,showpacs,floatfix
]{revtex4-1}

\usepackage{tabularx}
\usepackage{float}
\usepackage{bm}
\usepackage{graphicx}
\usepackage{amsmath}
\usepackage{mathtools}
\usepackage{color}
\usepackage{mathrsfs}
\usepackage[colorlinks=true,citecolor=blue]{hyperref}
\hypersetup{colorlinks=true,citecolor=blue,linkcolor=blue,urlcolor=blue}

\usepackage{chngcntr}

\begin{document}

\title{Tunable optical nonlinearity for TMD polaritons dressed by a Fermi sea}

\author{V. Shahnazaryan}
\affiliation{Department of Physics and Engineering, ITMO University, St. Petersburg, 197101, Russia}
\affiliation{Institute of Physics, Polish Academy of Sciences, Al. Lotnikow 32/46, 02-668 Warsaw, Poland}

\author{V. K. Kozin}
\affiliation{Science Institute, University of Iceland, Dunhagi-3, IS-107 Reykjavik, Iceland}
\affiliation{Department of Physics and Engineering, ITMO University, St. Petersburg, 197101, Russia}

\author{I. A. Shelykh}
\affiliation{Department of Physics and Engineering, ITMO University, St. Petersburg, 197101, Russia}
\affiliation{Science Institute, University of Iceland, Dunhagi-3, IS-107 Reykjavik, Iceland}

\author{I. V. Iorsh}
\affiliation{Department of Physics and Engineering, ITMO University, St. Petersburg, 197101, Russia}

\author{O. Kyriienko}
\affiliation{Department of Physics and Astronomy, University of Exeter, Stocker Road, Exeter EX4 4QL, UK}

\begin{abstract}
We study a system of a transition metal dichalcogenide (TMD) monolayer placed in an optical resonator, where the strong light-matter coupling between excitons and photons is achieved. We present quantitative theory of the nonlinear optical response for exciton-polaritons for the case of a doped TMD monolayer, and analyze in detail two sources of nonlinearity. The first nonlinear response contribution stems from the Coulomb exchange interaction between excitons. The second contribution comes from the reduction of Rabi splitting that originates from phase space filling at increased exciton concentration and the composite nature of excitons. We demonstrate that both nonlinear contributions are enhanced in the presence of free electrons. As free electron concentration can be routinely controlled by an externally applied gate voltage, this opens a way of electrical tuning of the nonlinear optical response.
\end{abstract}

\maketitle

\section{Introduction}

Planar microcavities in the regime of strong light-matter coupling represent a robust platform for nonlinear optics. The hybridization of excitons with optical cavity photons leads to the formation of ultralight interacting quasiparticles---polaritons---that enable polariton lasing \cite{Kasprzak2006,Balili2007,Schneider2013,Ballarini2017} and emergent polariton fluid behavior \cite{Amo2009,Amo2011,Tercas2014,CarusottoCiutiRev}. For conventional quantum well (QW) nanostructures in III-V and II-VI semiconductors in this regime various nonlinear effects were studied, including formation of solitons \cite{Sich2012,Hivet2012,Chana2015,Opala2018}, vortices \cite{Lagoudakis2008,Tosi2012,Boulier2015,Kwon2019,Caputo2019}, polarization multistability \cite{Gippius2007,Cerna2013,Gavrilov2013,Kyriienko2014,Klaas2017,Tan2018}, and nontrivial polariton lattice dynamics \cite{Askitopoulos2013,Ohadi2017,Sigurdsson2017,Gao2018,Sigurdsson2019,Mietki2018,Kyriienko2019a}.
These unique properties can be used for experimental realization of ultra fast  polariton-based nonlinear optical integrated devices \cite{ShelykhReview,Amo2010,Liew2010,Ohadi2015,Dreismann2016,Askitopoulos2018,Opala2019}.

In these systems the nonlinear response mainly originates from the Coulomb-based exciton-exciton scattering \cite{Ciuti1998,Tassone1999,Glazov2009,Vladimirova2010,Brichkin2011,Estrecho2019,Levinsen2019}, typically observed at macroscopic mode occupations. For high quality samples prerequisite signatures of a quantum nonlinear behavior were recently observed \cite{Munoz-Matunano2019,Delteil2019}, thanks to outstanding fabrication advances. The limitations for QW-based platform come from low operation temperatures, relatively small light-matter coupling ($\sim 4$~meV per QW), and complex growth techniques \cite{DengRev}.

Recent advances in the field of optically-active two-dimensional (2D) materials have largely increased capabilities of polaritonics \cite{Schneider2018,Mortensen2020}. In this case excitons are hosted by monolayers of transition metal dichalcogenide (TMD) materials---atom-thick nanostructures with a direct optical bandgap \cite{Mak2010,Chernikov2014,Miwa2015,Steinleitner2017,Rostami2015,Schwarz2014} and excellent optical properties \cite{Wang2018,Wurstbauer2017}. To date light-matter coupling for TMD excitons was observed in various configurations, including optical microcavities \cite{LiuMenon2015,Dufferwiel2015,Dufferwiel2017,Sidler2017,Emmanuele2019}, Tamm plasmon structures \cite{Lundt2017}, photonic crystals \cite{Zhang2018,Kravtsov2020}, surface plasmons \cite{Kleemann2017,Goncalves2018,Geisler2019} and nanoantennas \cite{Antosiewicz2014,Stuhrenberg2018}. Due to the relatively large electron/hole masses and reduced screening, TMD exciton binding energy ranges in hundreds of meVs, and multicharge bound complexes (trions \cite{Mak2013,Ross2013,Singh2016,Courtade2017,Lundt2018}, biexcitons \cite{You2015}) can be observed. Importantly, small exciton volume leads to large Rabi frequency, and excitonic optical response dominates already at room temperature \cite{Wang2018}. Further list of exceptional properties of TMD monolayers includes strong spin-orbit interaction and valley-dependent physics \cite{Wang2017a,Manca2017,Lundt2019}, peculiar exciton transport properties \cite{Kulig2018,Zipfel2020}, and strong dependence on dielectric properties for observed physical effects \cite{Huser2013,Latini2015,Shahnazaryan2019}. For doped and gated TMD samples bandgap renormalization was shown \cite{Chernikov2015a,Chernikov2015,Withers2015,Raja2017}, which opens the way to engineering of material properties. Finally, studies of trapped excitons and strain-induced lattices revealed efficient defect-based single photon emission from two-dimensional materials \cite{Kumar2016,Branny2017,Carmen2017,Flatten2018}.

The natural next step in TMD polaritons is exploring of the nonlinear response. This so far has proven to be a nontrivial task, as the very same large binding leads to reduction of the exciton-exciton scattering cross-section evidenced theoretically \cite{Shahnazaryan2017} and experimentally \cite{Barachati2018}.
However, the situation changes drastically once large light-matter coupling is achieved and a TMD monolayer is doped with free carriers. First, the deviation of excitonic statistics from ideal bosons \cite{CombescotReview,Combescot2006} leads to the nonlinear Rabi splitting behavior\cite{Brichkin2011,Daskalakis2014,Yagafarov2020,Betzold2020}---optical saturation---that in the case of strongly-coupled TMD polaritons was shown to give significant contribution \cite{Emmanuele2019}. Second, the presence of free electron gas (Fermi sea) strongly modifies the optical response of TMD monolayers. It depends on the density of the electrons, and leads to several characteristic regimes \cite{Chang2018,Shiau2017}. At low free electron concentrations sharp additional peak appears that is typically attributed to charged exciton complexes (trions), being bound states of two electrons and one hole \cite{Mak2013,Courtade2017,Emmanuele2019}. At high electron concentrations the broad spectral peak was observed and attributed to an exciton polaron-polariton---correlated state of an exciton dressed by the Fermi sea \cite{Efimkin17,Sidler2017,Ravets2018,Tan2020}. In each case the enhancement of the nonlinear response was reported \cite{Emmanuele2019,Tan2020,Kyriienko2019}.

One should note, however, that besides formation of the additional peak, corresponding to the appearance of new quasiparticles, the presence of free electrons shall modify substantially the optical response of the exciton mode itself. This is especially pronounced in the case of intermediate electron densities, where excitons are spectrally separated from the other modes. Differently from the cases of very low and high electron densities, this region remains unexplored so far. In the current paper we aim to bride the gap between the regimes of low and high electron densities, focusing on nonlinear optical properties of the system. In particular, we demonstrate that Fermi sea strongly contributes to the screening of Coulomb potential and the onset of additional correlations stemming from the Pauli exclusion principle. These effects change both light matter coupling and exciton-exciton interactions, thus resulting in renormalization of the strength of nonlinearity. Our theory shows that the change of the free electron density, which can be routinely realized in gated TMD samples, gives a powerful tool for controlling the optical nonlinearity.

The paper is organized as follows. In Section II we present the theory describing the behavior of an exciton in a TMD monolayer in the presence of Fermi sea. It accounts for both screening (S) of the Coulomb potential and the Pauli blocking (PB) effect. We study the modification of the excitonic wavefunctions, excitonic binding energy and Bohr radius.
In Section III we calculate exciton-exciton interaction potentials in the presence of an electron gas. We demonstrate that the screening and the Pauli blocking play opposite roles, the former increasing and the latter decreasing effective exciton-exciton interaction constant. We find that this counterintuitive effect of the Pauli blocking comes from the mixing of exciton ground and excited states.
As a result of the competition of these two mechanisms we report an overall increase of the interaction constant with electron density. In Section IV we analyze the impact of free electrons on nonlinear reduction of the Rabi splitting in the system, and show that nonlinearity increases with electron density. Section V summarizes our findings.


\section{2D excitons in the presence of a Fermi sea}
\begin{figure}[t!]
    \centering
    \includegraphics[width=0.95\linewidth]{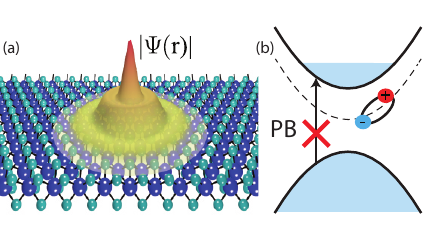}
    \caption{Sketch of the system. (a) Absolute value of an excitonic wavefunction in a TMD monolayer that is strongly modified by the free electron gas in the conduction band. (b) Sketch of the excitation process where the Pauli principle excludes occupied states in the conduction band preventing the exciton formation.
    }
    \label{fig:sketch}
\end{figure}

We study a transition metal dichalcogenide monolayer where optical response is strongly dominated by tightly-bound neutral Wannier excitons. Considering a doped monolayer, we account for the presence of the Fermi sea formed by the excessive charge. This can lead to the modification of an optical response in several different ways, and dominant contributions depend on the free electron density $n$ and correspondingly the location of the Fermi level $E_F$. At low electron concentrations charged excitons---trions---are formed. In TMD monolayers these three-particle bound states have binding energy $E^T_b$ being much smaller than an exciton binding energy $E^X_b$. Therefore, in the low-density regime $E_F\ll E^T_b$ trion- and exciton-based response is spectrally well-separated. However, properties of excitons in doped monolayers are modified by an electron gas through screening and Pauli blocking (Fig.~\ref{fig:sketch}).
In the high density regime where $E_F\sim E^T_b$ strong many-body correlations between excitons and electrons become important, and the system is described in terms of exciton-polarons~\cite{Efimkin17} (dressed exciton-electron quasiparticles). For instance, in MoS$_2$ monolayer with $E^T_b = 18$~meV this corresponds to the concentration of excess carriers $n\sim 10^{12}.. 10^{13}$~cm$^{-2}$.

In the present work we focus on the low- and intermediate-density regime, where the exciton-based optical response is modified by the Fermi sea through the exciton wavefunction and energy renormalization. To account for electrons we solve the Wannier equation for the eigenenergy $E_{\mathrm{X}}$ and the momentum-space exciton wavefunction $C_{\mathbf{p}}$ that reads\cite{Efimkin17}
\begin{equation}
\label{eq:Wannier}
    \bigg(\frac{\hbar^2 k^2}{2 \mu}+\Sigma_g\bigg)C_{\mathbf{k}}-\sum_{\mathbf{k'}} B_{\mathbf{k}}V_{\mathbf{k}-\mathbf{k'}}B_{\mathbf{k'}}C_{\mathbf{k'}}=E_{\mathrm{X}} C_{\mathbf{k}},
\end{equation}
%
where the exciton binding energy $E^X_b = |E_{\mathrm{X}}-\Sigma_g|$ accounts for the band gap renormalization $\Sigma_g$ caused by excessive charge carriers. In Eq.~\eqref{eq:Wannier} $\mu=m m_v/(m+m_v)$ is an exciton reduced mass, and $m$ and $m_v$ stand for the conduction and valence band effective mass, respectively. $B_{\mathbf{k}}=[1-n_F(E^c_{\mathbf{k}})]^{1/2}$ is the Pauli blocking  factor that excludes filled electronic states from the space available for exciton formation, $E^c_{\mathbf{k}}$ denotes an energy dispersion for the conduction band, and $n_F$ is a Fermi-Dirac distribution. To account for the effects of screening caused by the excess charge carriers~\cite{Pol2DEG,Glazov2018} and the atomic thickness of the material~\cite{KeldyshRytova}, we consider the screened interaction potential
\begin{equation}
V_{\mathbf{k}}=\frac{2\pi  e^2}{(4\pi\varepsilon_0\kappa)[k+\rho_0 k^2+k_{sc}(k)]},
\end{equation}
where $\varepsilon_0$ is the vacuum permittivity, $\rho_0$ is a screening parameter associated to the intrinsic polarizability of the two-dimensional layer, $k_{sc}(k)=-2\pi  e^2\Pi(k)/(4\pi\varepsilon_0\kappa)$ is the screening momentum, and $\kappa$ denotes a dielectric constant of the surrounding media. We use the static polarization operator of two-dimensional electron gas~\cite{Pol2DEG}  $\Pi(k)=-m/(\pi\hbar^2)[1-\Theta(k-2k_F)(1-4k_F^2/k^2)^{1/2}]$, where the Fermi wavevector is $k_F=\sqrt{2 m E_F}/\hbar$. $\Sigma_g$ accounts for the bandgap renormalization by carriers due to screening and phase space filling effects, and reads
\begin{equation}
    \Sigma_g=-\sum_{\mathbf{k}}V_\mathbf{k}n_F(E^c_\mathbf{k})-\sum_{\mathbf{k}}(V^0_\mathbf{k}-V_\mathbf{k})n_F(E^v_\mathbf{k}),
\end{equation}
where $E^c_\mathbf{k}=\hbar^2 \mathbf{k}^2/2m-E_F$ and $E^v_\mathbf{k}=\hbar^2 \mathbf{k}^2/2m_v-E_g-E_F$ denote the energies of conduction and valence bands, $E_g$ is the non-screened bandgap width and $V^0_\mathbf{k}=2\pi  e^2/[4\pi\varepsilon_0\kappa (k+\rho_0 k^2)]$.
For the sake of simplicity we neglect the renormalization of electron masses in conduction and valence bands, and retain only the bandgap renormalization  $\Sigma_g$.

The rotational symmetry of the potential $V_{\mathbf{k}}$ allows one to write the wavefunction in the form
\begin{equation}
    C_{\mathbf{k}}=C^{n,m_z}(k)\frac{e^{i m_z \theta}}{\sqrt{2\pi}}
\end{equation}
where we use polar coordinates $\mathbf{k}=(k,\theta)$. Then the Wannier equation \eqref{eq:Wannier} for $C^{n,m_z}(k)$ reads
\begin{align}
    \label{eq: exciton_Schrodinger}
    &E_{\mathrm{X}} C^{n,m_z}(k)= \bigg(\frac{\hbar^2k^2}{2\mu}+\Sigma_g\bigg)C^{n,m_z}(k)+\\
    &\int_0^{\infty} \frac{k'dk'}{(2\pi)^2} B(k)V_1(k,k')B(k')C^{n,m_z}(k')\nonumber
\end{align}
where $V_1(k,k')$ is
\begin{align}
&V_1(k,k')= \nonumber \\
&-\int_0^{2\pi} d\theta V(\sqrt{k^2+k^{'2}-2kk'\cos{\theta}})e^{i m_z\theta}.
\end{align}

\begin{figure}
    \centering
    \includegraphics[width=0.98\linewidth]{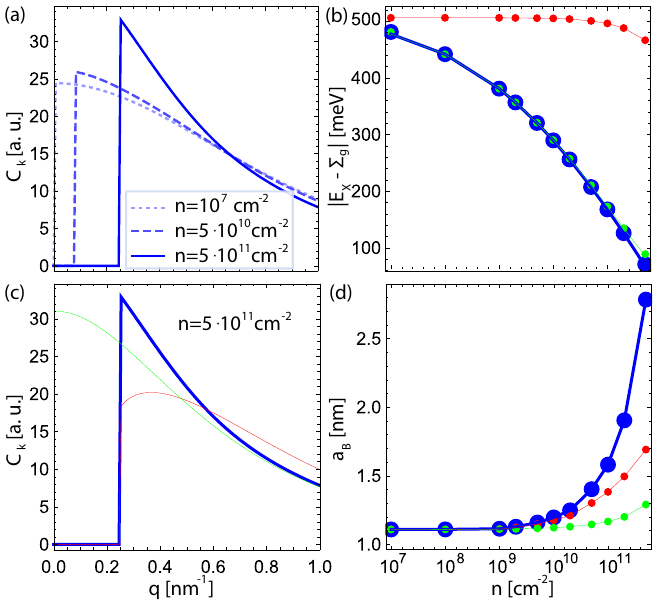}
    \caption{ (a) Exciton wavefunction in the momentum space shown for different electron gas density. The shift of wavefunction maximum is caused by the increase of Fermi energy and the corresponding wavevector.
    (b) Exciton binding energy as a function of free electron concentration. Here the green curve corresponds to the absence of Pauli blocking, the red curve corresponds to the absence of interaction screening by electron gas, and the blue solid curve accounts both effects.
    (c) The impact of screening and Pauli blocking factors on exciton wavefunction at high density of free electron gas. Colors are the same as in panels (b). Notably, screening by the free electron gas leads to rescaling of hydrogen-like wave function (green curve), whereas the Pauli blocking determines the modified shape of the wavefunction (red curve).
    (d) Bohr radius shown as a function of free electron concentration. Labelling is the same as in (b).
 }
    \label{fig:WF}
\end{figure}

We solve Eq.~\eqref{eq: exciton_Schrodinger} numerically for a monolayer of transition metal dichalcogenide. The structure parameters vary a lot throughout the  literature and depend on the choice of both TMD monolayer material and its surrounding. Here we set the screening length to $r_0=4$ nm and the bandgap to $E_g=2.6$ eV, which are typical for MoS$_2$ layer \cite{Stier2018,Chernikov2014,Qiu2013}. We also fix equal effective masses, $m_v=m$. As a reference, we choose the case of freestanding monolayer ($\kappa=1$), and set $m=0.35m_0$ ($m_0$ is a free electron mass), being typically the case for TMD monolayers \cite{Larentis2018}.
We consider only the exciton ground-state, so that we set $m_z=0$.
The results of calculations are shown in Fig.~\ref{fig:WF}. In Fig.~\ref{fig:WF}(a) we present the momentum space distribution for the excitonic wavefunction. We observe that the increase of free electron gas density leads to the strong modification of the wavefunction as compared to standard two-dimensional hydrogen-like wavefunction that has a form $\propto \left[ 1+ (\lambda k)^2 \right]^{-3/2}$. The quenching of low-momenta region stems from the Pauli blocking effect.
In order to extract the relative contributions of Pauli blocking factor and the screening of interaction induced by electron gas, we simulate Eq.~\eqref{eq: exciton_Schrodinger} in the regimes when one of the factors is effectively turned off. At low density the impact of both effects is small, and the wavefunctions plotted for this two cases nearly coincide (not shown). At relatively high density of the free electron gas the momentum space wavefunction is shown in Fig.~\ref{fig:WF}(c) (thick blue curve). We see that the additional screening leads to re-scaling of wavefunction [Fig.~\ref{fig:WF}(c), green curve], while the Pauli blocking is responsible for the suppression of low momenta region [Fig.~\ref{fig:WF}(c), red curve].

Plots in Fig.~\ref{fig:WF}(b, d) illustrate the electron density dependence of the exciton binding energy and Bohr radius. The latter is defined as an average electron-hole separation $a_B(n) =\langle \psi_{n}(r) |r|\psi_{n}(r) \rangle$, where $\psi_{n}(r)$ is the exciton wavefunction in the real space, and we highlight that it depends on the free electron gas density $n$. The growth of the electron concentration leads to stronger interaction screening, which results in weaker binding of excitons. In the absence of screening the Pauli blocking factor becomes essential for larger values of electron density, leading to reduction of binding energy and corresponding increase for the exciton Bohr radius.
%


\section{Exciton-exciton interaction}

Next, we study the exciton-exciton interaction processes for TMD monolayers that originate from Coulomb interaction of electrons and holes. We use the standard scattering theory approach \cite{Tassone1999,Ciuti1998,Glazov2009} and exploit the calculated exciton wavefunction to account for the presence of the electron gas. First, we note that the direct interaction is suppressed due to the electron-hole equal effective masses, $m_v=m$ \cite{Ciuti1998}. Hence, the total interaction constant $g_{\mathrm{tot}}$ is determined by the electron and hole exchange terms, which are identical due to equal effective masses. Thus, $g_{\mathrm{tot}}=2g^e_{\mathrm{exch}}$, with $g^e_{\mathrm{exch}}$ denoting the electron exchange interaction constant. The latter reads \cite{Glazov2009}
\begin{align}
    g^e_{\mathrm{exch}} (Q) =\frac{2}{A}\sum_{\mathbf{k},\mathbf{q} } &V_{\mathbf{q}}
    C_{ \mathbf{k}-\mathbf{q}/2 }
    C_{ \mathbf{k}-(\mathbf{Q}-\mathbf{q})/2 }
    C_{\mathbf{k}+\mathbf{q}/2 }
    \notag \\
    &
    \left[
    C_{ \mathbf{k}+(\mathbf{Q}-\mathbf{q})/2 } -
    C_{ \mathbf{k}+(\mathbf{Q}+\mathbf{q})/2 }
    \right],
\end{align}
where $\mathbf{Q}$ is an exchanged momentum, and $A$ is the normalization area.

\begin{figure}
    \centering
    \includegraphics[width=0.98\linewidth]{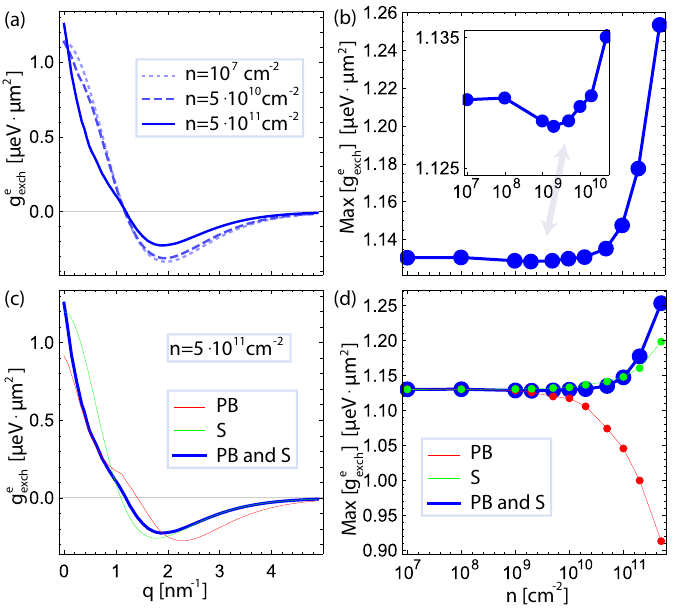}
    \caption{Exciton-exciton exchange interaction. (a) The dependence of interaction constant on transfer momenta at different densities of free electron gas. While the interaction maxima demonstrates a moderate and non-monotonous shift with the increase of the free electron gas density, the shape for transfer momenta dependence is nearly unaltered.
    (b) The maxima of interaction constant as a function of free electron gas density. The inset illustrates zoomed-in region with the non-monotonous dependence.
    (c) The impact of screening and Pauli blocking factors on momentum dependence of exciton-exciton interaction at high density of free electron gas. Here the green curve corresponds to the absence of Pauli blocking, the red curve to the absence of interaction screening by electron gas, and blue solid curve accounts both effects.
    (d) The influence of screening and Pauli blocking factors on the maxima of exchange interaction vs the density of free electrons. Colors are the same as in panel (c). }
    \label{fig:exchscr}
\end{figure}

In Fig.~\ref{fig:exchscr}(a) we present the dependence of the electron exchange interaction constant as a function of exchange momenta. The maximum of interaction constant demonstrates a moderate increase with increasing $n$, and the shape of exchange momenta dependence is generally unchanged. Fig.~\ref{fig:exchscr}(b) presents the dependence of interaction maxima on the electron density. Particularly one can see that the dependence is non-monotonous, with the local minima appearing at moderate electron densities. The latter stems from complex interplay between multiple factors, discussed below.

In order to understand the origin of this non-trivial dependence of interaction on the electron gas density, we perform calculations (i) in the absence of Pauli blocking factor, and (ii) in the effective absence of screening. The corresponding dependence on exchange momenta is shown in Fig.~\ref{fig:exchscr}(c) for $n=5\cdot 10^{11}$~cm$^{-2}$ density of free electrons.
We observe that the interaction has its highest value when both effects are accounted (blue thick curve).
In the absence of Pauli blocking the screening leads to the slight decrease of interaction maxima [case (i), green curve]. In turn the Pauli blocking leads to significant reduction [case (ii), red curve].
In Fig.~\ref{fig:exchscr}(d) we plot the maxima of interaction versus the
density of the electron gas. We observe that screening leads to the monotonous enhancement of interaction [case (i), green curve], and the Pauli blocking leads to its monotonous reduction [case (ii), red curve].

The enhancement of the interaction coefficient due to the interaction screening [case (i)] is caused by the enhancement of exciton Bohr radius that dominates the weakening of interaction potential. The origin of reduction due to the Pauli blocking [case (ii)] stems from the fact that the Pauli blocking leads to mixing of exciton ground and excited states. In its turn, it was shown earlier, that the interaction between excited exciton states is of attractive nature \cite{Shahnazaryan2016,Shahnazaryan2017}, explaining the overall decrease of repulsive interaction between excitons. Here we find that the exciton wave function in the presence of Pauli blocking can be expanded in terms of $1s$, $2s$, $3s$ exciton states, and the calculation of exciton-exciton interaction in terms of such functions agrees well with the one calculated in the presence of Pauli blocking (see Fig.~\ref{fig:expansion}). The details of the calculation are shown in Appendix \ref{app:A}.
The presence of both Pauli blocking and screening leads to a complicated dependence on the density of free electron gas, with regions dominated by the reduction stemming from Pauli blocking and the enhancement arising from screening, as depicted in Fig.~\ref{fig:exchscr}(d)].

\begin{figure}
    \centering
    \includegraphics[width=0.98\linewidth]{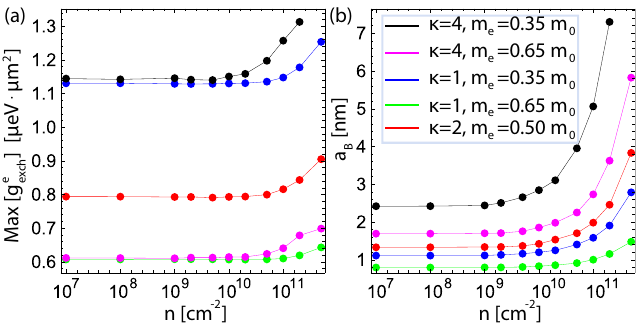}
    \caption{(a) The maximum of exciton-exciton exchange interaction and (b) exciton Bohr radius shown as a function of the free electron gas density for different material parameters. Colors in both panels correspond to parameters shown in panel (b). The increase of the dielectric constant $\kappa$ for surrounding media leads to the growth of exciton Bohr radius, which is nearly compensated by the reduction of interaction potential between excitons. Instead, the reduction of effective mass leads to the increase of Bohr radius, which is not compensated by change in interaction potential. This shows that for the fixed effective mass the interaction constant has weak dependence on the dielectric environment properties.}
    \label{fig:exchpar}
\end{figure}

We further analyze the dependence of exciton-exciton interaction on material and substrate parameters. The results are shown in Fig.~\ref{fig:exchpar}. We observe that the interaction constant demonstrates non-monotonous dependence with local minima at intermediate density regardless the structure parameters. The latter means that the observed effect is of general character and does not depend qualitatively on the material choice. It is remarkable that the growth of Bohr radius due to increase of dielectric constant is nearly compensated by the corresponding reduction of interaction potential, leading to overall weak dependence of exciton-exciton constant on the dielectric properties of surrounding media [cf. blue and black curves in Fig.~\ref{fig:exchpar}(a)]. On the other hand, the smaller effective mass of electrons leads to larger Bohr radius, resulting in enhancement of exchange interaction, as the increase of Bohr radius here is not compensated by corresponding reduction of interaction potential [cf. blue and green curves in Fig.~\ref{fig:exchpar}(a)].
It should be mentioned, that all the characteristic peculiarities of the free electron gas impact on the exciton  Coulomb nonlinearity remain unaltered in the case of unequal effective masses of conduction and valence bands (see Appendix \ref{app:B} for the details).

%


\section{Saturation effects and quenching of the Rabi splitting}

\begin{figure}
    \centering
    \includegraphics[width=0.98\linewidth]{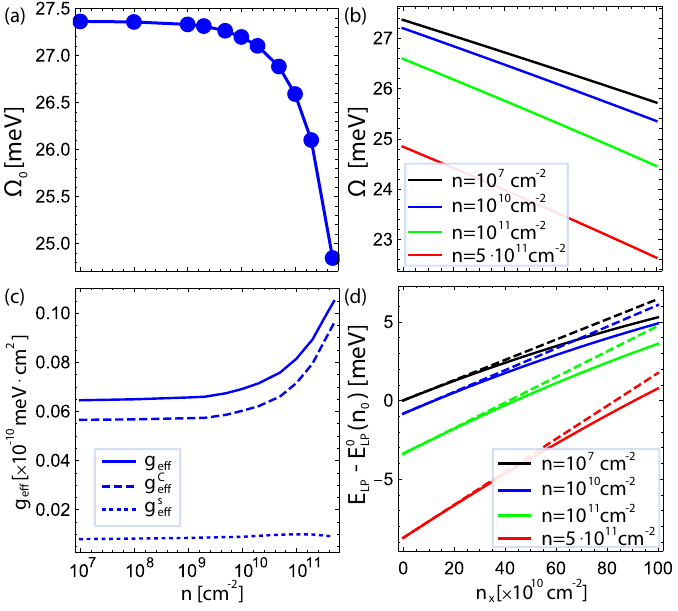}
    \caption{(a) Rabi splitting as a function of the free electron density in the weak excitation regime.
    (b) Light-matter coupling as a function of exciton density shown for different electron gas concentration.
    (c) Polariton nonlinearity coefficient and its contributions shown as a function of the free electron density.
    (d) Energy of the lower polariton branch relative to the reference value $E^0_{\mathrm{LP}} (n_0)$ as a function of the exciton density at different $n$. Here the solid curves correspond to Eq.~\eqref{eq:E_LP}, and the dashed curves to Eq.~\eqref{eq:E_LPe}.
    }
    \label{fig:Rabi}
\end{figure}

We proceed with the discussion of impact of free electron gas on the coupling between exciton and cavity photon modes. For a TMD monolayer put in a microcavity this corresponds to the electron density-dependent Rabi frequency. It can be expressed as
\begin{equation}
    \label{eq:coupling}
    \Omega_0 (n)  = \sqrt{ \frac{E_C}{\kappa \varepsilon_0 L_C} } |\psi_n (0)| d_{cv},
\end{equation}
where $E_C$ is the cavity resonance energy, $L_C=\pi \hbar c /(\sqrt{\kappa}E_C)$ is the cavity length, and $c$ is the speed of light. Here $\psi_n (0)$ is the real space exciton wavefunction at the origin that depends on the free electron gas density.
We note that the latter cannot be approximated via conventional relation $\psi_n (0) \propto a_B^{-1}$ due to the Pauli blocking, leading to the mixing of exciton ground and excited states.
Finally, $d_{cv}$ denotes the dipole matrix element for the optical interband transition.

We consider the case of the optical cavity being resonant to the exciton transition in the absence of electron gas, $E_C=E_{\mathrm{X}}^0(0)$. The exciton transition energy in the presence of electron gas reads $E_{\mathrm{X}}^0 (n)=E_g +\Sigma_g (n)- E_b(n)$, where we recall that $E_b=|E_{\mathrm{X}} -\Sigma_g|$ is the exciton binding energy.
It should be noted, that the position of excitonic transition varies very slowly with the increase of $n$, as the reduction of binding energy is largely compensated by the corresponding bandgap renormalization. The latter is in good agreement with experimental \cite{Chernikov2015,Ugeda2014,Lin2014} and theoretical evidence  \cite{Shahnazaryan2019}.

The value of the dipole matrix element of interband transition is set to $d_{cv}=7$ D, leading to Rabi splitting of $\sim 30$ meV, in agreement with existing experimental results \cite{Dufferwiel2017}. Here we assume that the dependence of the interband transition matrix element on the density of free electron gas is negligible. Hence, the density of electron gas affects on the efficiency light-matter coupling only via the exciton wave function [see Eq.~\eqref{eq:coupling}]. The dependence of light-matter coupling $\Omega_0(n)$ on the density of free electron gas is presented in Fig.~\ref{fig:Rabi}(a). The reduction of coupling with the increase of electron density stems from the impact of Pauli blocking, and the detailed analysis is presented in Appendix \ref{app:C}.

Next we study the nonlinear part of light-matter interaction that is represented by optical saturation coming from the phase space filling. Recently it was shown to provide a significant nonlinear response contribution for TMD polaritons \cite{Emmanuele2019, Kyriienko2019}. Together with nonlinear exciton-exciton interaction, the optical saturation effect leads to the energy blueshift for the lower polariton mode, coming from the renormalization of Rabi splitting. It depends on the density of excitons $n_{\mathrm{X}}$ and the excitonic wavefunction. The generalized Rabi frequency can be written as \cite{Yagafarov2020}
\begin{align}
\label{eq:Omega_N}
\Omega (n_{\mathrm{X}},n) \approx \Omega_0 (n) \sqrt{1 - 2 s(n) n_{\mathrm{X}}} ,
\end{align}
where the saturation factor
\begin{equation}
    s(n)=\frac{\sum_{\mathbf{k}} |C_{\mathbf{k}}|^2 C_{\mathbf{k}}}{\sum_{\mathbf{k'}} C_{\mathbf{k'}}^*}
\end{equation}
accounts for the phase space filling arising from multiple exchange diagrams.
In particular, in the case of effectively hydrogenic wavefunctions, this yields $s^{\mathrm{hyd}}={8\pi  a_B^2}/7$, meaning that the larger Bohr radius provides larger nonlinearity. Here, however, the presence of Pauli blocking leads to a moderate dependence of the saturation factor on the density of free electron gas, discussed in Appendix \ref{app:C}. As stated in Eq.~\eqref{eq:Omega_N}, for growing density of excitons the Rabi splitting effectively shrinks. The corresponding dependence is illustrated in Fig.~\ref{fig:Rabi} (b) for various values of free electron gas density.

The energy of lower polariton branch reads as
\begin{align}
    \label{eq:E_LP}
    E_{\mathrm{LP}} (n_{\mathrm{X}},n) &= \frac{1}{2} \bigg[ E_C + E_{\mathrm{X}} (n,n_{\mathrm{X}})  \notag \\
    & -\sqrt{\left[E_C-E_{\mathrm{X}} (n,n_{\mathrm{X}})\right]^2 +\Omega^2(n_{\mathrm{X}},n)}  \bigg],
\end{align}
where the exciton energy is $E_{\mathrm{X}}(n,n_{\mathrm{X}})=E_{\mathrm{X}}^0(n)+ g_{\mathrm{tot}(n)} n_{\mathrm{X}} /2$.
Introducing the detuning between cavity and exciton modes as $\Delta(n)=E_C-E_{\mathrm{X}}(n)$, and taking the limit of low exciton density,
the energy for the lower polariton mode reads
\begin{equation}
    \label{eq:E_LPe}
    E_{\mathrm{LP}} (n_{\mathrm{X}}, n)  \approx E_{\mathrm{LP}}^0 (n)+ g_{\mathrm{eff}}(n) n_{\mathrm{X}},
\end{equation}
which consists of the linear part equal to
\begin{equation}
    \label{eq:E_LP0}
    E_{\mathrm{LP}}^0 (n)= E_C- \frac{\Delta+\sqrt{\Delta^2+\Omega_0^2} }{2},
\end{equation}
and nonlinear blueshift $g_{\mathrm{eff}}(n) n_{\mathrm{X}}$. Here $g_{\mathrm{eff}}(n)$ is an effective polariton nonlinearity coefficient that is a sum of Coulomb-based interaction and saturative nonlinearity contributions,
\begin{align}
    g_{\mathrm{eff}}(n) &= \left( 1+ \frac{\Delta}{\sqrt{\Omega^2_0 +\Delta^2} } \right) \frac{g_{\mathrm{tot}} }{4}
    + \frac{\Omega^2_0 }{2\sqrt{\Omega^2_0 +\Delta^2} }s \notag \\
     &=:g^C_{\mathrm{eff}}(n)+g^s_{\mathrm{eff}}(n).
      \label{eq:g_eff}
\end{align}
For the compactness we omitted the electron density dependence of the quantities $\Delta$, $\Omega_0$, $s$ etc.

Polaritonic nonlinearity coefficient and its parts are shown in the Fig.~\ref{fig:Rabi}(c) as a function of free electron gas density. Notably we observe that while the saturation coefficient $s$ increases more than the Coulomb interaction, the electron density dependence of its pre-factor diminishes its enhancement, making it nearly flat (dotted curve). Instead, the increase of Coulomb nonlinearity is further boosted by the corresponding growth of its pre-factor (the dashed curve).

In Fig.~\ref{fig:Rabi}(d) the dependence of the lower polariton energy on the exciton density is shown, where we plot its nonlinear contribution (as compared to $E^0_{\mathrm{LP}} (n_0)$ with $n_0=10^7$ cm$^{-2}$). At fixed density of free electrons the increase of the exciton density leads to both reduction of light-matter interaction, and the blueshift of the exciton energy. In each this leads to the blueshift of the lower polariton energy.
With the increase of free electron gas density both the exciton-exciton interaction constant $g^e_{\mathrm{exch}}$ and the saturation factor $s$ are enhanced, leading to the corresponding growth of the nonlinear optical response.
It should be mentioned that relation $E_{\mathrm{LP}} (n_{\mathrm{X}}, n)  \approx E_{\mathrm{LP}}^0 (n)+ g_{\mathrm{eff}}(n) n_{\mathrm{X}}$ is valid in the moderate excitation regime $n_{\mathrm{X}} \leq 10^{12}$ cm$^{-2}$, as for higher intensities the quadratic terms $\propto n_{\mathrm{X}}^2$ become relevant.

Finally, we discuss the impact of trion on the exciton-polariton nonlinear optical response. As stated in Section II, the trion resonance is far detuned from the exciton and cavity modes, and thus has limited impact on resulting polariton modes at higher energy. Indeed, the estimate for the trion binding energy for the considered case of freestanding monolayer is about 50 meV, and it mixing with polariton branches is small (see Appendix \ref{app:D} for the details). The detailed analysis of the electron density dependence of trion nonlinear response for the case of near-resonant cavity is a separate research question, and will be studied in future works.

\section{Conclusions}

In this paper we analyzed the behavior of exciton polaritons in a TMD monolayer in the presence of a gas of free electrons. We revealed that the Fermi sea has a strong effect on the nonlinear optical response of the system. We found that the role of free electrons is twofold.
First, doping leads to screening of the Coulomb interaction, and results in the increase of exciton Bohr radius and simultaneously the reduction of exciton-exciton interaction potential. Our calculations show that the overall impact of the screening leads to the enhancement of exciton-exciton interaction coefficient.
Second, due to the Pauli exclusion principle, the presence of the free electrons also dramatically modifies the structure of the excitonic wave functions, suppressing the contribution of the harmonics corresponding to small electron wavevectors. Surprisingly the impact of the Pauli blocking factor leads to the reduction of exciton-exciton interaction. We found that the latter can be attributed to mixing of exciton ground and excited states, caused by the Pauli blocking factor. It is known that the interaction between excited exciton states is of attractive type, which explains the reduction of exciton-exciton repulsive interaction caused by Pauli principle. Finally, we showed that the combined impact of interaction screening and Pauli blocking leads to the non-monotonous dependence of the exciton-exciton interaction constant as a function of free electron gas density.

The presence of Fermi sea substantially modifies also the statistics-based renormalization of the Rabi splitting at high exciton densities, which gives another contribution to the enhancement of the optical nonlinearity.
It is important to note that both Coulomb nonlinearity and saturation-based nonlinearity generally grow with the increase of the free electron gas density.
As the  latter can be easily controlled by application of the external gate voltage, our findings pave the way to accessible and experimental friendly tuning of the degree of optical nonlinearity in TMD based samples.

\section*{acknowledgments}

The authors are grateful to D. Efimkin for valuable discussions.
This work was supported by the Russian Science Foundation (grant No. 19-72-00171). O.K. thanks the support from UK EPSRC New Investigator Award under the agreement number EP/V00171X/1. V.K.K., I.A.S. and O.K. acknowledge support from Icelandic Science Foundation project ``Hybrid polaritonics''.\\


\appendix
\counterwithin{figure}{section}

\section{Expansion of an exciton wavefunction in terms of basis functions}\label{app:A}

\begin{figure}[h!]
    \centering
    \includegraphics[width=0.95\linewidth]{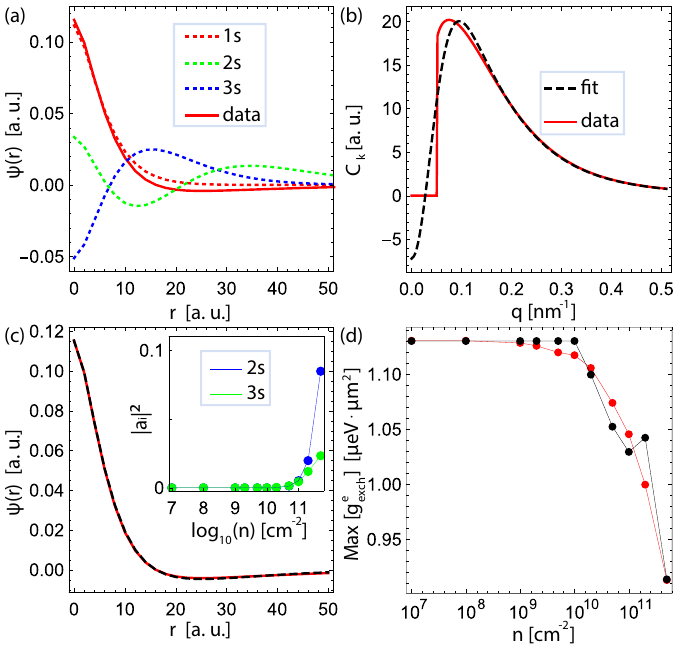}
    \caption{ (a) The real space dependence of exciton wave function in the presence of Pauli blocking at density of free electrons $n=5 \cdot 10^{11}$ cm$^{-2}$ (red solid curve), and the exciton ground and excited states wavefunctions in its absence. Here we neglect the impact of the interaction screening by free electrons.
    (b) Real and (c) momentum space dependence of wave function (red solid curve) and its expansion in terms of basis functions (black dashed curve).
    The inset in panel (c) demonstrates the contribution of excited states versus the density of free electron gas.
    (d) The maxima of exciton-exciton interactions as a function of free electron density. The red dots correspond to calculation using the actual wave functions, and the black dots is calculated using the wavefunctions expanded in terms of basis functions. The mismatch in the values is attributed to the imperfection of the fitting procedure.
    }
    \label{fig:expansion}
\end{figure}

We analyze the impact of Pauli blocking factor on the excitonic wavefunction that results in the reduction of the exciton-exciton interaction for increasing electron gas density. First, we simulate Eq.~\eqref{eq: exciton_Schrodinger} of the main text in the absence of both interaction screening and the Pauli blocking. We find wavefunctions for ground and excited exciton states ($1s$, $2s$, $3s$). Their real space distributions are shown in Fig.~\ref{fig:expansion}(a). Next, for each value of $n$ we expand calculated wavefunctions in the presence of Pauli blocking (but non screening) in terms of bare basis functions. This procedure yields
\begin{align}
    \label{eq:expansion}
    \psi_n(r) = a_1 (n) \psi_{1s} (r)
    +a_2 (n) \psi_{2s} (r)
    +a_3 (n) \psi_{3s} (r),
\end{align}
where coefficients $a_i(n)$  are found from
\begin{align}
    & \psi_n(0) =a_1 (n) \psi_{1s} (0)
    +a_2 (n) \psi_{2s} (0)
    +a_3 (n) \psi_{3s} (0), \notag \\
    & \psi_n(r_1) =a_1 (n) \psi_{1s} (r_1)
    +a_2 (n) \psi_{2s} (r_1)
    +a_3 (n) \psi_{3s} (r_1), \notag \\
    & 1=|a_1(n)|^2+|a_2(n)|^2+|a_3(n)|^2,
\end{align}
and $r_1$ corresponds to the first root of $\psi_n$. The results of fitting for large density of the electron gas are shown in Fig.~\ref{fig:expansion}(b, c). We find that the fit is nearly exact at small $r$, but strongly deviates at large distances (not shown). Correspondingly, for the small momenta there is a strong deviation, while for larger values there is a good agreement. Finally, in Fig.~\ref{fig:expansion}(d) we provide the comparison of the exciton-exciton interaction coefficient calculated from exact wave functions (red dots), and the wave functions defined by the \eqref{eq:expansion}. Evidently, with the increase of electron gas density Pauli blocking leads to larger contribution of excited states [see the inset in Fig.~\ref{fig:expansion}(c)], which interact attractively \cite{Shahnazaryan2016,Shahnazaryan2017}, resulting in corresponding reduction of the ground state exciton-exciton repulsive interaction.

\section{The dependence of Rabi splitting saturation rate on the free electron gas density}\label{app:B}
\begin{figure}
    \centering
    \includegraphics[width=0.95\linewidth]{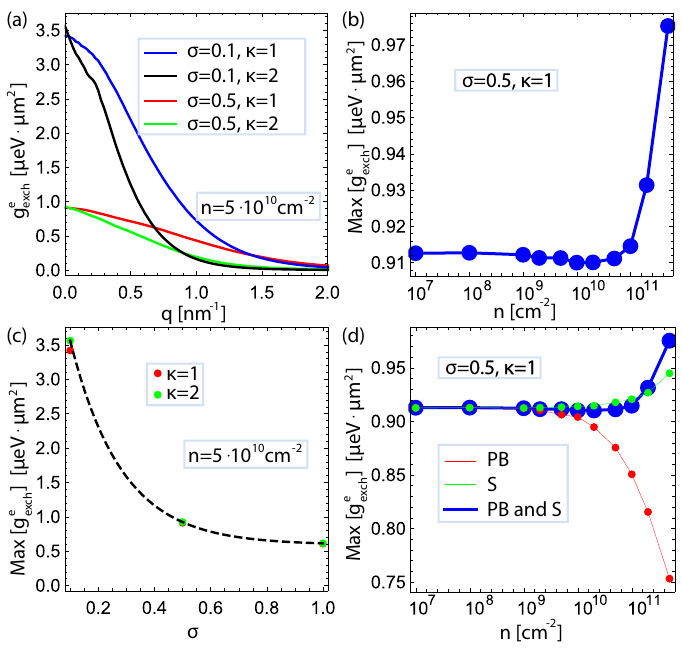}
    \caption{ Exciton-exciton exchange interaction. (a) The dependence of interaction constant on transfer momenta for different dielectric screening and effective mass ratio $\sigma$.
    (b) The maxima of interaction constant as a function of free electron  density.
    (c) The effective mass ratio dependence of exchange interaction  constant. Dashed black curve corresponds to an exponential fit.
    (d) The influence of screening and Pauli blocking factors on the maxima of exchange interaction vs the density of free electrons. }
    \label{fig:sigma}
\end{figure}

The effective masses of conduction and valence bands in TMD monolayer typically differ from each other \cite{Kidd2016}.
Here we study the impact of mass ratio $\sigma=m/m_v$ on the exciton-exciton interaction strength. To do so, we fix the effective mass of hole as $m_v=0.65m_0$, and consider the cases when $\sigma = 0.5$, $\sigma =0.1$. We also consider two values of surrounding dielectric constant, $\kappa=1$ and $\kappa=2$. The results of calculations are shown in Fig.~\ref{fig:sigma}. It is evident, that the absolute values of interaction are strongly dependent on the effective masses, and weakly dependent on the dielectric screening [Fig.~\ref{fig:sigma} (a)]. Moreover, the maximum of interaction constant demonstrates an exponential behavior on the effective mass ratio $\sigma$ [Fig.~\ref{fig:sigma} (c)]. However, the impact of Pauli blocking and the interaction screening by free electrons together with the nonmonotonous dependence of interaction on the free electron gas density are qualitatively independent on the structure parameters [Fig.~\ref{fig:sigma} (b) and (d)].

\section{The dependence of Rabi splitting saturation rate on the free electron gas density}\label{app:C}

\begin{figure}
    \centering
    \includegraphics[width=0.95\linewidth]{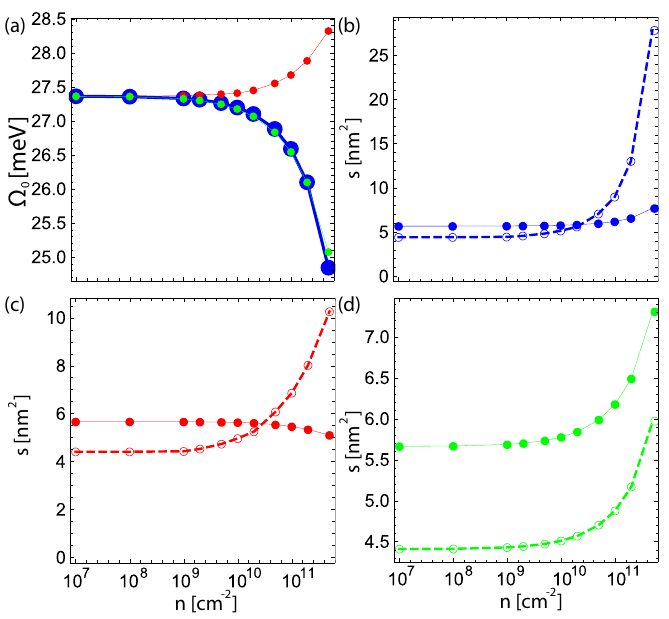}
    \caption{ (a) Light-matter coupling as a function of the free electron gas density at low excitation regime.  Here green curves correspond to the absence of Pauli blocking, red curves correspond to the absence of interaction screening, and blue curves account both effects.
    (b), (c), (d) The saturation factor $s(n)$ (solid curves) and its hydrogen-like estimate $s^{\mathrm{hyd}} (n)$  (dashed curves) versus the free electron gas density.
    Panel (b) corresponds to the presence of both Pauli blocking and the interaction screening;
    panel (c) stands for the absence of the interaction screening;
    panel (d) illustrates the absence of Pauli blocking.
    }
    \label{fig:saturation}
\end{figure}

In Fig.~\ref{fig:saturation}(a) we present light-matter coupling at small exciton densities, $n_{\mathrm{X}} a_B^2 \ll 1$, as a function of free electron gas density $n$. In the absence of the Pauli blocking the screening of Coulomb interaction leads to weaker binding of excitons, so that the wavefunction is less concentrated around the origin. The latter results in the quenching of light-matter coupling [Fig.~\ref{fig:saturation}(a), green curve]. On the contrary, in the absence of screening the Pauli blocking leads to mixing with excited exciton states, leading to the increase of wavefunction amplitude at the origin [see Fig.~\ref{fig:expansion}(a)]. This results in the corresponding enhancement of light-matter coupling [Fig.~\ref{fig:saturation}(a), red curve]. Yet, the impact of screening is much stronger, so that the interplay of this counteracting effects leads to overall reduction of light-matter coupling with the increase of free electron gas density [Fig.~\ref{fig:saturation}(a), blue curve].

We further analyze the Rabi splitting saturation factor $s(n)$. Fig.~\ref{fig:saturation}(b) illustrates its dependence on the density of free electron gas.
We observe a moderate enhancement of the saturation rate with the growth of the electron gas density. On the other hand, the estimate of saturation factor for the hydrogen-like exciton $s^{\mathrm{hyd}} (n)$ grows much faster  [dashed curve in Fig.~\ref{fig:saturation}(b)].
To get a better insight, we study the impacts of the screening of interaction and Pauli blocking separately.

In Fig.~\ref{fig:saturation}(c) we present the free electron gas density dependence of saturation factor and its estimate in the absence of interaction screening. As the density increases, the saturation factor moderately reduces, while its estimate enhances.
This discrepancy with the hydrogen-based estimate stems from the Pauli blocking, which leads to emergence of strongly non-hydrogenic wavefunctions.

Fig.~\ref{fig:saturation}(d) illustrates the case of the absence of Pauli blocking and the presence of interaction screening. Here both the saturation factor and its estimate increase nearly on equal footing, indicating that in the absence of Pauli blocking effect the hydrogen-like model is valid up to a constant.
In total, the interplay of these two counteracting impacts results in the moderate enhancement of saturation efficiency, as depicted in Fig.~\ref{fig:saturation}(b).

\section{The impact of trion state on the exciton-polariton spectra}\label{app:D}

\begin{figure}
    \centering
    \includegraphics[width=0.95\linewidth]{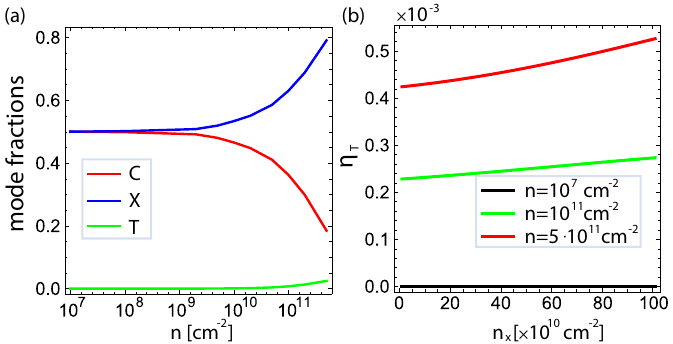}
    \caption{ (a) The fractions of photon (C), exciton (X) and trion (T) modes in the middle polariton branch as a function of free electron gas density.
    (b) Correction of middle polariton branch energy by trions, plotted as a function of the exciton density.
}
    \label{fig:TPol}
\end{figure}

The binding energy of trion can be estimated using the variational energy minimization with Chandrasekhar-type trial function in momentum space \cite{Ramon2003, Emmanuele2019}
\begin{align}
    \Phi^\mathrm{T}_{k_1,k_2} = \mathcal{N} \left[
    \Phi_{k_1}(\lambda_1) \Phi_{k_2}(\lambda_2) +
    \Phi_{k_2}(\lambda_1) \Phi_{k_1}(\lambda_2)
    \right],
\end{align}
where $\Phi_{k_i}(\lambda_j) = \sqrt{8\pi \lambda_j^2}\left[  1+ (\lambda_j k_i)^2 \right]^{-3/2}$, $\mathcal{N}=\left[ 2(1+\chi^2) \right]^{-1/2}$ and $\chi = 4 \lambda_1 \lambda_2 / (\lambda_1 + \lambda_2)^2$.
The results for the considered parameters of freestanding monolayer are $E_b^{\mathrm{T}} = 52.83$ meV, $\lambda_1 = 1.21$~nm, $\lambda_2 = 3.06$~nm.  For an estimate we neglect the screening of Coulomb interactions by free electron gas, meaning that the position of trion resonance remains unaltered when the free electron density changes.
Though this treatment is simplistic, it allows to get in the first approximation the impact of trions on exciton-polariton spectra.

The contribution of the trion resonance can be analysed within the model of three coupled modes, where together with dominant exciton-photon coupling there is an admixture of the far-detuned trion mode. Polariton eigenmodes can be obtained diagonalizing
\begin{align}
    \hat{H}^\mathrm{T} = \begin{pmatrix}
    E_\mathrm{C} & \Omega_\mathrm{X}(n,n_\mathrm{X}) & \Omega_\mathrm{T}(n,n_\mathrm{X}) \\
    \Omega_\mathrm{X}(n,n_\mathrm{X}) & E_\mathrm{X}(n,n_\mathrm{X}) & 0 \\
    \Omega_\mathrm{T}(n,n_\mathrm{X}) & 0 & E_\mathrm{T}
    \end{pmatrix} ,
    \label{eq:HamTP}
\end{align}
where $E_\mathrm{T} = E_\mathrm{C} - E_b^\mathrm{T}$, and the trion-photon coupling is calculated as \cite{Emmanuele2019}
\begin{equation}
    \Omega_\mathrm{T} (n,n_\mathrm{X}) = 4\mathcal{N}\left( \frac{\lambda_1}{\lambda_2} +\frac{\lambda_2}{\lambda_1} \right)
    \frac{\sqrt{n}}{|\psi_n(0)|} \Omega_\mathrm{X} (n,n_\mathrm{X}).
\end{equation}
The eigenmodes of Hamiltonian \eqref{eq:HamTP} correspond to three polariton branches. Here we focus on the middle polariton branch $E_{\mathrm{MP}}^\mathrm{T}$, as it is dominated by exciton mode with a small admixture of trion, as depicted in Fig.~\ref{fig:TPol}(a). As the electron density increases, the trion contribution slowly increases due to reduction of detuning and the enhancement of trion-photon coupling. Fig.~\ref{fig:TPol}(b) demonstrates the exciton density dependence of spectrum correction caused by trions. The latter is defined as
\begin{equation}
    \eta_\mathrm{T} = \frac{E_{\mathrm{MP}}^\mathrm{T} - E_{\mathrm{LP}} }{E_{\mathrm{LP}} } .
\end{equation}
We observe that even for large concentration of free electrons the correction is minor and thus can be neglected in the description of the middle polariton branch.



\end{document}